\documentclass[a4paper]{spie}  
\emergencystretch=\maxdimen 
\usepackage{amsmath,amsfonts,amssymb}
\usepackage{graphicx}
\usepackage[colorlinks=true, allcolors=blue]{hyperref}

\title{Assessing ionospheric effects on image quality in the LOFAR Decametre Sky Survey}

\author[a,b]{C.M. Cordun}
\author[a,b]{H.K. Vedantham}
\author[a]{M. Mevius}
\affil[a]{ASTRON, Netherlands Institute for Radio Astronomy, Oude Hoogeveensedijk 4, Dwingeloo 7991PD, The Netherlands}
\affil[b]{Kapteyn Astronomical Institute, University of Groningen, PO Box 800, 9700 AV Groningen, The Netherlands}

\authorinfo{Further author information: (Send correspondence to C.M.C)\\C.M.C.: E-mail: cordun@astron.nl}

\begin{document} 
\maketitle

\begin{abstract}
The terrestrial ionosphere is a major limitation for high angular resolution imaging at decametre wavelengths. Its rapidly varying and direction-dependent phase errors can defocus sources, reduce recovered flux densities, and produce strong field-to-field variations in image quality. We investigate these effects using three overlapping pointings from the LOFAR Decametre Sky Survey (LoDeSS), observed under different ionospheric conditions. Using phase-calibration solutions on a bright source, we derive the spatial diffractive scale, the temporal decorrelation time, and the angular scale over which direction-dependent calibration solutions can be transferred before residual phase errors become significant. These metrics are reflected in the final image quality. The field with the weakest ionospheric phase structure has the sharpest compact sources and the most reliable flux-density scale, while the field with the strongest phase structure shows substantial source broadening and flux-density loss. The comparison indicates that spatial ionospheric structure is the dominant limitation in these observations. Calibration-derived ionospheric metrics, therefore, provide a practical way to diagnose and predict image-quality variations in decametre survey data.
\end{abstract}


\section{INTRODUCTION}
\label{sec:intro}  

The decametre sky ($\sim$10--30~MHz) remains largely unexplored at sub-arcminute resolutions. In this frequency range, source spectra can differ strongly from their behaviour at metre and centimetre wavelengths because absorption, spectral ageing, and low-energy particle populations become increasingly important \cite{condon1992radio,callingham2015broadband}. As a result, decametre observations are useful for identifying steep-spectrum and curved-spectrum emission from radio galaxies and cluster-scale diffuse emission.  \cite{2008A&ARv..15...67M,brunetti2020second,mandal2020revived,shulevski2019first,tabatabaei2017radio,2025Natur.647..603C,driessen2018investigating,stappers2011observing}. 
The decametre band is also expected to contain the coveted cyclotron emission from Jupiter-like exoplanets that is unavailable at higher frequencies \cite{zarka1998auroral,zarka2007plasma,griessmeier2007predicting}. However, observations at these frequencies have been limited by the difficulty of producing reliable interferometric images through the rapidly varying terrestrial ionosphere.

Decametre imaging is affected by three main observational and calibration challenges: (a) radio-frequency interference that is pernicious below 30\,MHz, (b) contamination from bright sources \cite{offringa2013lofar,blaszkiewicz2021comparative}  that are invariably present in the large primary fields of view at these frequencies, and (c) the terrestrial ionosphere. The first two effects can be partially mitigated by observing at night and advanced flagging and demixing, respectively \cite{offringa2012morphological,van2007self}. The ionosphere is less straightforward. Its corruption of interferometric phase varies with time, observing frequency, station position, and direction on the sky. The problem is severe at decametric wavelengths due to the proximity of the observing band to the ionospheric plasma frequency, which means that small changes in electron content can introduce large phase changes, and that the higher-order terms must be used when the phase is modelled as a function of frequency  \cite{de2018effect}. If these phase errors are only partially corrected, they reduce coherence on long baselines, leading to defocused sources with artificially suppressed peak brightness and errors in the flux-density scale.

For automated imaging necessary in large surveys, the variable nature of the ionosphere is particularly problematic. Its structure changes from observation to observation, cannot be forecast with sufficient accuracy for calibration, and may vary on angular and temporal scales comparable to or smaller than those sampled by the calibration scheme. As a result, observations with the same instrumental setup and processing strategy can produce substantially different image quality when taken under different ionospheric conditions.

Catalogues and maps from instruments such as UTR-2, Clark Lake, DRAO, and the 8C survey provided much of the existing view of the sky at tens of MHz \cite{braude1978decametric,kassim1988clark,roger1999radio,rees1990deep}. These data remain important, but they were obtained with much coarser angular resolution and lower sensitivity than is achievable with modern low-frequency arrays. Recent work has shown that sub-arcminute imaging at decametre wavelengths is possible in selected fields \cite{osinga2024probing,groeneveld2024characterization,groeneveld2025serendipitous}. However, the conditions under which such imaging succeeds or fails are still not fully quantified. In particular, it remains necessary to connect measurable properties of the ionospheric phase screen to the final image products.

The Low Frequency Array (LOFAR \cite{VanHaarlem2013}) provides a suitable data set for this test through the LOFAR Decametre Sky Survey (LoDeSS \cite{cordun_lodess_submitted}). LoDeSS uses the LOFAR Low Band Antenna system to image the northern sky above a declination of $20^\circ$ over the frequency range 15--30\,MHz, with an observing time of approximately 5~h per pointing. In this work, we select three LoDeSS pointings that overlap on the sky but were observed under different ionospheric conditions. This configuration allows us to isolate the residual effects of the ionosphere post-calibration. Our goal is to determine whether ionospheric metrics measured during calibration can explain the observed field-to-field differences in decametre image quality. To do so, we derive spatial and temporal ionospheric structure functions from the calibrator phase solutions and use them to quantify the expected image-quality degradation.

This paper is structured as follows. Section~\ref{sec:data_proc} describes the selected LoDeSS observations and the calibration and imaging procedure. The ionospheric metrics used in this work are defined in Section~\ref{sec:iono_analysis} and applied to the three pointings in Section~\ref{sec:results}. Section~\ref{sec:discussion} compares these metrics with the image quality and flux-density recovery, and Section~\ref{sec:conclusion} draws the main conclusions.

\section{OBSERVATIONS AND DATA PROCESSING}
\label{sec:data_proc}

The data analysed in this work are taken from Data Release 1 of LoDeSS.
We selected three partially overlapping pointings: P139+79, P222+74, and P264+86. Their basic observational properties are listed in Table~\ref{tab:selected_pointings}. 

\begin{table}[hb]
\centering
\begin{tabular}{lcccccc}
\hline\hline
Field & RA [deg] & Dec [deg] & Obs. duration [h] & Flagged [\%] & RMS [mJy~beam$^{-1}$] \\
\hline
P139+79 & 139.091 & 79.052 & 5   & 23.4 & 12.58 \\
P222+74 & 222.379 & 74.494 & 6   & 20.8 & 11.46 \\
P264+86 & 264.991 & 86.543 & 8  & 14.5 & 8.79  \\
\hline
\end{tabular}
\caption{Observation details of the selected fields.}\label{tab:selected_pointings}
\end{table}

The data were processed with a modified version of the Library for Low Frequencies (LiLF) \cite{2019A&A...622A...5D,2020A&A...642A..85D}, a calibration and imaging pipeline developed for LOFAR LBA data. We used the LiLF version \url{https://github.com/revoltek/LiLF}, commit \texttt{d5f4104}. The full LoDeSS processing strategy, including the modifications made for the decametre regime, is described in the accompanying LoDeSS DR1 paper \cite{cordun_lodess_submitted}. Here, we summarise only the main steps and the products used in this work.

The processing consists of four main stages: pre-processing, primary calibrator calibration, transfer of the calibrator solutions to the target fields, and direction-dependent self-calibration of the target field. During pre-processing, radio-frequency interference (RFI) is flagged, bright off-axis sources are demixed, and the data are averaged to 24.4\,kHz in frequency and 4\,s in time. We then use the primary calibrator pipeline to derive the polarisation alignment, phase, and bandpass solutions. These solutions are transferred to the target field before we perform direction-dependent self-calibration and imaging.

Several modifications to the standard LiLF processing were required for the 15--30\,MHz LoDeSS data. In the calibrator pipeline, we flatten the bandpass before applying an additional flagging step over the full observing band. The flat spectral baseline post bandpass correction allows us to more effectively identify broadband radio-frequency interference, which we find to be particularly important below 30\,MHz. In the target-field self-calibration, we reduced the time and frequency smoothing of the solutions so that the calibration could better track the rapid temporal and spectral changes at decametre wavelengths. The frequency-smoothness constraint was therefore set to $\sim1\,\mathrm{MHz}$ for the outer remote stations and to $\sim5\,\mathrm{MHz}$ for the core stations. The corresponding gain solutions were obtained on $15\,\mathrm{min}$ intervals for the core stations and on $1$--$4\,\mathrm{min}$ intervals for the remote stations, with the shortest timescales used for the outermost stations. For fields containing very bright sources, we use a modified calibration strategy in which the calibration is first restricted to the bright-source subfield before the full field is calibrated. This reduces artefacts caused by direction-dependent errors and ionospheric scintillation.

The final products used in this paper are the primary-beam-corrected Stokes-I images of the overlapped regions and the calibrator phase solutions. The images are used to quantify the final effect of ionospheric phase corruption. The calibrator phase solutions are used to quantify the ionospheric metrics described in Section~\ref{sec:iono_analysis}.

\section{IONOSPHERIC METRICS}
\label{sec:iono_analysis}
For the ionospheric analysis, we use the phase solutions derived on the bright primary calibrator. These phases are dominated by two contributions: instrumental clock delays and ionospheric total electron content (TEC). We separate these terms using the improved clock--ionosphere separation method developed for LOFAR data \cite{cordun2025improved}, and retain only the ionospheric TEC component for the following analysis. The method models the ionospheric phase to second order\cite{appleton1932wireless,cordun2025improved}:
\begin{equation}
\begin{split}
    &\phi_{i,{\rm iono}}(\tau_{1,2},\nu)= -\frac{2\pi\nu}{c} \int_{\rm LoS} (n-1) \,{\rm d}l \approx \frac{2\pi}{c}\left(\frac{a}{\nu}\tau_1+\frac{a^2}{2\nu^3}\tau_2\right),\\
    & {\rm where} \ \ \ a = 10^{16}\frac{e^2}{8\pi^2\epsilon_0m_e},\qquad {\rm and} \ \ \ \tau_n = \int_{\rm LoS} N_e^n \,{\rm d}l .
\end{split}
\label{eq:tec_to_phase}
\end{equation}
Here, $\phi_{i,{\rm iono}}$ is the ionospheric phase for station $i$, $\nu$ is the observing frequency, $c$ is the speed of light, $e$ is the elementary charge, $m_e$ is the electron's mass, $\epsilon_0$ is the vacuum permittivity, $n$ is the ionospheric refractive index, and ${\rm d}l$ is the line element along the line of sight (${\rm LoS}$). The first order term in the parentheses contains $\tau_1$ which is the line-of-sight TEC, expressed in TECU ($1\,{\rm TECU}=10^{16}\,{\rm electrons}\,{\rm m}^{-2}$). Although the clock--ionosphere separation includes the second-order term (with $\tau_2$), we only include the first-order TEC term to calculate our ionospheric metrics. 

Before computing the ionospheric metrics, we subtract the median TEC value for each station within each time chunk. This removes global TEC offsets, leaving only the variations relevant to the structure-function analysis. We analyse the observations in one-hour chunks because it matches the contiguous observing blocks used in LoDeSS.

\subsection{Diffractive scale}
Our first metric is the characteristic spatial scale over which the ionospheric phase varies, called the diffractive scale. We estimate it by constructing the spatial phase structure function. At spatial offset $r$, this function is given by the variance of the phase difference between two locations separated by distance $r$. The definition implicitly assumes statistical isotropy on the two-dimensional phase screen above the telescope. Empirically, we compute it as the temporal variance of the ionospheric phase difference between two stations separated by $r$. That is, for each station pair $(i,j)$, the estimate is
\begin{equation}
D_\phi(r_{i,j},\nu)=\left\langle\left[\phi_{i,{\rm iono}}(\tau_{1,2},\nu)-\phi_{j,{\rm iono}}(\tau_{1,2},\nu)\right]^2\right\rangle_t .
\label{eq:spatial_sf}
\end{equation}

We assume a Kolmogorov-like phase structure function,
\begin{equation}
D_\phi(r,\nu)=\left(\frac{r}{r_{\rm diff}(\nu)}\right)^{5/3},
\label{eq:rdiff_model}
\end{equation}
The diffractive scale is therefore the distance at which the phase variance reaches 1 rad$^2$. Smaller values of $r_{\rm diff}$ correspond to stronger ionospheric phase variations.

Note that the diffractive scale, $r_{\rm diff}$, is frequency dependent since the ionospheric phase is frequency dependent. For the first-order TEC term in Eq.~\ref{eq:tec_to_phase}, the phase scales as $\phi_{\rm TEC}\propto\nu^{-1}$. Since the structure function is a variance, its amplitude scales as $\nu^{-2}$. Writing Eq.~\ref{eq:rdiff_model} as $D_\phi(r,\nu)=A(\nu)r^{5/3}$ gives $A(\nu)=r_{\rm diff}^{-5/3}$, and therefore $r_{\rm diff}\propto\nu^{6/5}$. 

\subsection{Decorrelation time}
Our second metric is the phase decorrelation timescale at a given location. It is defined analogously to the spatial diffractive scale using the phase variations at each station. For station $i$, the temporal structure function is
\begin{equation}
D_{\phi,i}(\Delta t,\nu)=\left\langle\left[\phi_{i,{\rm iono}}(\tau_{1,2},\nu,t+\Delta t)-\phi_{i,{\rm iono}}(\tau_{1,2},\nu,t)\right]^2\right\rangle_t ,
\label{eq:temporal_sf}
\end{equation}
where $\Delta t$ is the time lag. We fit the temporal structure function with
\begin{equation}
D_{\phi,i}(\Delta t,\nu)=\left(\frac{\Delta t}{t_{\rm decorr,i}(\nu)}\right)^{5/3},
\label{eq:tdiff_model}
\end{equation}
where $t_{\rm decorr,i}(\nu)$ is the station-dependent decorrelation time, defined as the lag at which the phase variance reaches 1 rad$^2$. Shorter $t_{\rm decorr,i}(\nu)$ values, therefore, correspond to faster ionospheric variability. 

If the temporal variations are caused by lateral advection of an unchanging phase screen at speed $v$, then $t_{\rm decorr}\sim r_{\rm diff}/v$ and is expected to follow the same frequency scaling, $t_{\rm decorr}\propto\nu^{6/5}$. Characteristic night-time advection speeds are around $100\,{\rm m}\,{\rm s}^{-1}$ which corresponds to $3600\,{\rm km}$ of drift over the one-hour time chunks. This spatial drift is much larger than anticipated outer scales of ionospheric fluctuations of around $100\,{\rm km}$. Equations \ref{eq:tdiff_model} and \ref{eq:rdiff_model} are only valid below the outer scale, at which point the structure function growth saturates. For this reason, we fit only the initial rising part of the structure function. We manually picked the 'rising part' for each dataset after perusing their structure function.

To estimate a single representative $t_{\rm decorr}$ for a given field and hour-long time chunk, we first take the median over stations within each chunk. Because the fitted quantities enter the structure functions through the amplitudes $A_r=r_{\rm diff}^{-5/3}$ and $A_t=t_{\rm decorr}^{-5/3}$, we average these amplitudes rather than the derived scales directly. This avoids giving disproportionate weight to quieter time intervals with large $r_{\rm diff}$ or $t_{\rm decorr}$. After averaging, we convert the weighted amplitudes back to effective values of $r_{\rm diff}$ and $t_{\rm decorr}$.

\subsection{Angular decorrelation scale}
\begin{figure}[!ht]
    \centering
    \begin{minipage}{0.49\linewidth}
        \centering
        \includegraphics[width=\linewidth]{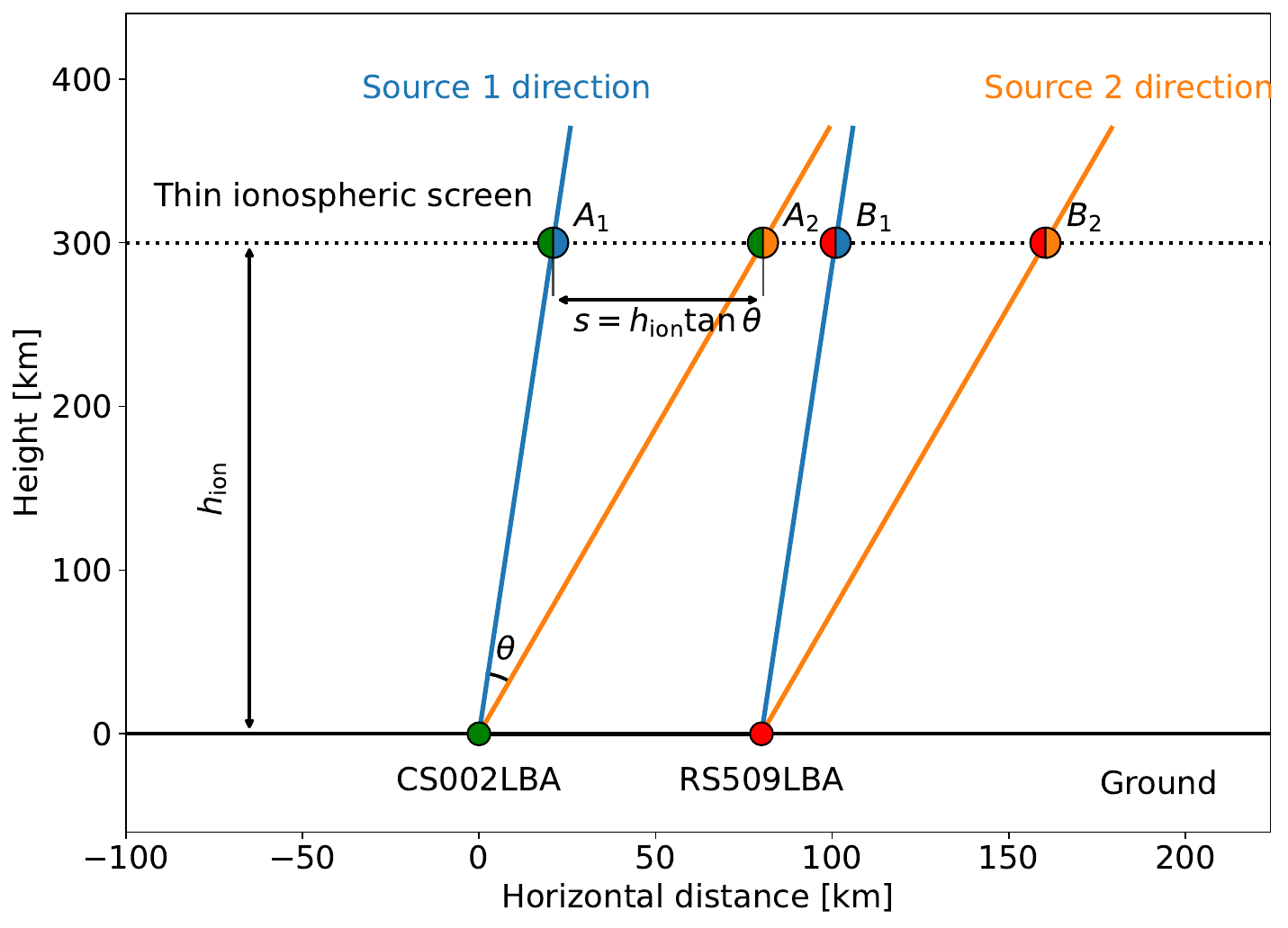}
        {(a) \small Side view of the ionospheric screen}
    \end{minipage}
    \hfill
    \begin{minipage}{0.49\linewidth}
        \centering
        \includegraphics[width=\linewidth]{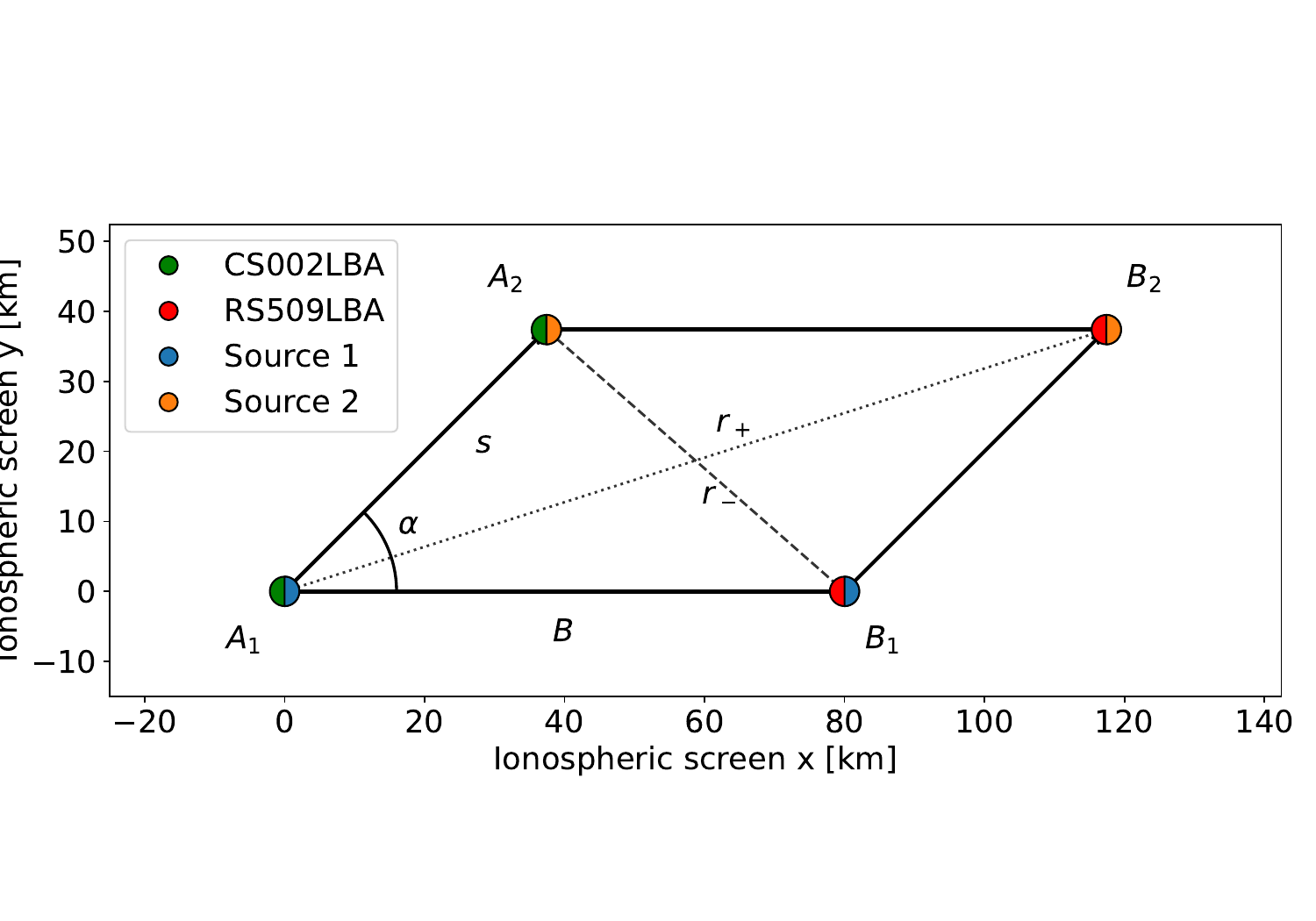}
        {(b) \small Top view of the ionospheric screen}
    \end{minipage}
    \vspace{0.1cm}
    \caption{Geometry used to estimate the angular scale for ionospheric calibration transfer.}
    \label{fig:iono_geometry}
\end{figure}

Finally, we estimate our third metric -- the angular separation over which calibration transfer becomes significantly affected by ionospheric phase differences. The geometry is illustrated in Fig.~\ref{fig:iono_geometry}. We adopt a plane-parallel approximation where the ionosphere is a thin phase screen at height $h_{\rm ion} = 300~km$. In the side view, two source directions are separated on the sky by an angle $\theta$. At the ionospheric screen, this angular separation corresponds to a physical separation
\begin{equation}
s(\theta) = h_{\rm ion}\tan\theta .
\label{eq:screen_separation}
\end{equation}
Thus, $s$ is the distance between the two pierce points produced by the two source directions for the same station.

The top-view geometry shows the same configuration projected onto the ionospheric screen. For source direction 1, the two stations pierce the screen at $A_1$ and $B_1$, while for source direction 2, the corresponding pierce points are $A_2$ and $B_2$. The separations $A_1B_1$ and $A_2B_2$ are both equal to the station baseline $B$. We use the maximum available baseline from the reference station, CS002LBA--RS509LBA, and denote this separation by $B_{\max}$. The separations $A_1A_2$ and $B_1B_2$ are equal to $s=h_{\rm ion}\tan\theta$, the pierce-point separation corresponding to an angular source separation $\theta$.

For a given source direction, the ionospheric contribution to the interferometric phase is the difference between the station phases. For source 1, this is
\begin{equation}
\Delta\phi_1 = \phi(B_1)-\phi(A_1),
\end{equation}
while for source 2 it is
\begin{equation}
\Delta\phi_2 = \phi(B_2)-\phi(A_2).
\end{equation}
If calibration solutions derived in direction 1 are transferred to direction 2, the residual phase error is the difference between these two baseline phases,
\begin{equation}
\delta\phi
=
\Delta\phi_2-\Delta\phi_1
=
\left[\phi(B_2)-\phi(A_2)\right]
-
\left[\phi(B_1)-\phi(A_1)\right].
\label{eq:transfer_error}
\end{equation}

We can estimate the variance of $\delta\phi$ using the measured spatial phase structure function. The structure function gives the expected phase variance between two points on the screen separated by a distance $r$,
\begin{equation}
D_\phi(r) =\left\langle\left[\phi(\mathbf{x}+\mathbf{r})-\phi(\mathbf{x})\right]^2\right\rangle .
\end{equation}
The four-point expression in Eq.~\ref{eq:transfer_error} can therefore be written in terms of the separations between the four pierce points. For a fixed orientation, this gives
\begin{equation}
\begin{split}
\sigma_\phi^2(\theta,\alpha)&=2D_\phi(B_{\max}) + 2D_\phi(s) - D_\phi(r_+) -D_\phi(r_-),\\
r_\pm&=\left(B_{\max}^2+s^2\pm 2B_{\max}s\cos\alpha \right)^{1/2}.
\end{split}
\label{eq:sky_angle_variance}
\end{equation}
Here, $\alpha$ is the angle between the baseline vector and the source-separation vector on the screen. The distances $r_+$ and $r_-$ are the two diagonal separations between the four pierce points, as shown in Fig.~\ref{fig:iono_geometry}b. The $D_\phi(B_{\max})$ term comes from the station separation for the two source directions, the $D_\phi(s)$ term comes from the separation between the source directions at the two stations, and the diagonal terms account for the cross-correlations between the two baseline phase differences.

For a fixed value of $\theta$, the result depends on the orientation angle $\alpha$. Since we want a single representative angular scale rather than a value for one particular orientation, we integrate over $\alpha$,
\begin{equation}
\left\langle \sigma_\phi^2(\theta) \right\rangle_\alpha=\frac{1}{\pi}\int_0^\pi\sigma_\phi^2(\theta,\alpha)\,{\rm d}\alpha .
\label{eq:sky_angle_orientation_average}
\end{equation}
The sky decorrelation angle $\theta_{\rm decorr}$ is then defined as the angular separation for which the orientation-averaged RMS transfer error reaches 1 rad,
\begin{equation}
\left[\left\langle \sigma_\phi^2(\theta_{\rm decorr}) \right\rangle_\alpha\right]^{1/2}=1~{\rm rad}.
\label{eq:theta_1rad}
\end{equation}

Although the structure function is derived from the TEC solutions in the calibrator field, the calibrator and target observations are taken simultaneously. We therefore assume that the ionospheric conditions above the two fields are sufficiently similar for this metric to apply to the target field. In this context, $\theta_{\rm decorr}$ provides an estimate of the angular distance over which solutions from a facet calibrator can be transferred before ionospheric residuals become large enough that a new facet calibrator is needed.

\section{RESULTS}
\label{sec:results}

We computed the ionospheric metrics described in Section~\ref{sec:iono_analysis}: (a) the spatial diffractive scale $r_{\rm diff}$, (b) the temporal decorrelation time $t_{\rm decorr}$, and (c) the sky separation at which the expected calibration-transfer error reaches 1 rad, $\theta_{\rm decorr}$, for the three selected LoDeSS pointings.  
The values for each field are presented in Table~\ref{tab:iono_results}.

\begin{table}[!ht]

\centering
\begin{tabular}{lcccccccc}
\hline\hline
Field & $r_{\rm diff}^{\rm 15MHz}$ & $r_{\rm diff}^{\rm 30MHz}$ & $t_{\rm decorr}^{\rm 15MHz}$ & $t_{\rm decorr}^{\rm 30MHz}$ & $\theta_{\rm decorr,CS302}^{\rm 15MHz}$ & $\theta_{\rm decorr,CS302}^{\rm 30MHz}$ & $\theta_{\rm decorr,RS509}^{\rm 15MHz}$& $\theta_{\rm decorr,RS509}^{\rm 30MHz}$ \\
      & [km] & [km] & [s] & [s] & [deg] & [deg] & [deg] & [deg] \\
\hline
P139+79 & 3.0 & 7.3 & 25.8 & 62.9 & $>$3 & $>$3 & 0.45 & 1.18\\
P222+74 & 0.6 & 1.4 & 6.0 & 14.5 & 0.11 & 0.36 & 0.08 & 0.20\\
P264+86 & 1.0 & 2.3 & 10.3 & 25.3 & 0.20 & 2.52 & 0.14 & 0.34\\
\hline
\end{tabular}
\caption{Summary of the ionospheric metrics for the three selected pointings.}\label{tab:iono_results}
\end{table}

\subsection{Spatial ionospheric structure}

\begin{figure}[!ht]
    \centering
    \begin{minipage}{0.49\linewidth}
        \centering
        \includegraphics[width=\linewidth]{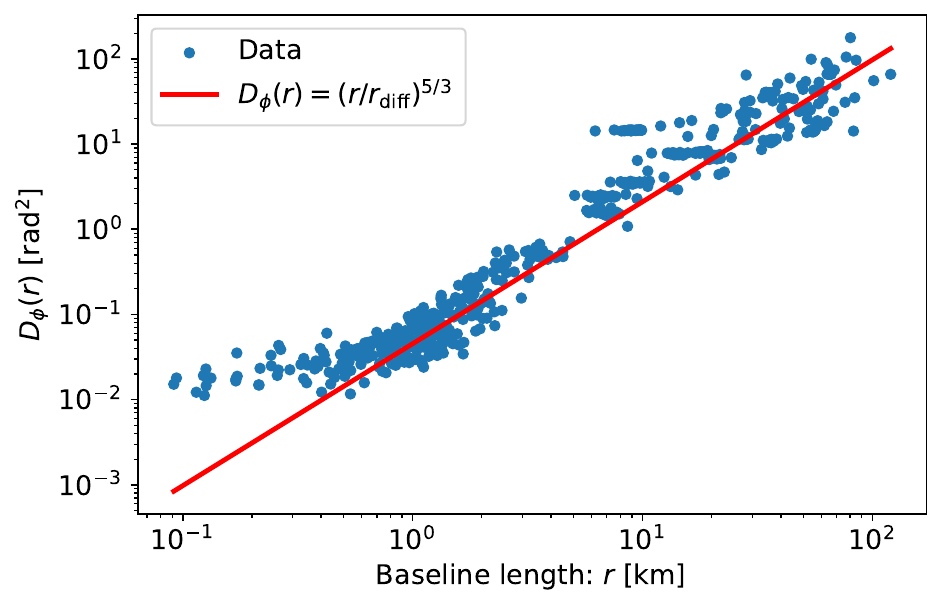}
        {(a) \small Example spatial phase structure function $D_\phi(r)$ and fitted Kolmogorov-like model for the first hour of P139+79 at 30~MHz.}
    \end{minipage}
    \hfill
    \begin{minipage}{0.49\linewidth}
        \centering
        \includegraphics[width=\linewidth]{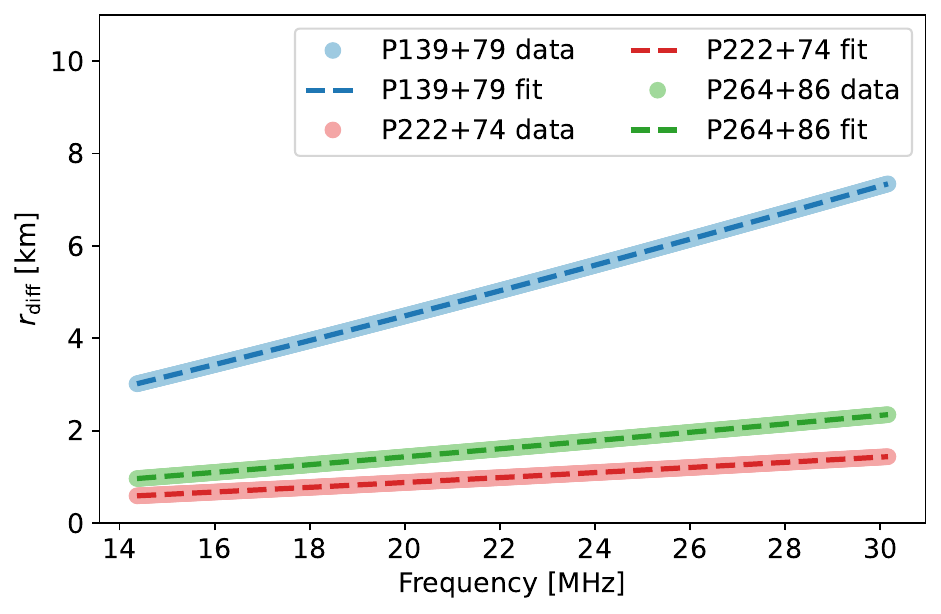}
        {(b) \small Frequency-dependent $r_{\rm diff}$ values for all three selected pointings. The fitted curves follow the $r_{\rm diff}\propto\nu^{6/5}$ scaling.}
    \end{minipage}
    \vspace{0.1cm}
    \caption{Spatial ionospheric structure-function analysis and fitted diffractive scales.}
    \label{fig:rdiff}
\end{figure}

Figure~\ref{fig:rdiff} shows the spatial ionospheric structure-function analysis. The example structure function at 30~MHz in Fig.~\ref{fig:rdiff}a follows the expected Kolmogorov-like form over the fitted baseline range of $\approx 0.4-100$\,km, allowing the diffractive scale $r_{\rm diff}$ to be estimated using Eq.~\ref{eq:rdiff_model}. The flattening of the measured structure function below $\approx 0.4$\,km is likely caused by a combination of thermal-noise fluctuations dominating the ionospheric fluctuations and baseline lengths falling below the Fresnel scale, where the pierce-point approximation breaks down \cite{2015MNRAS.453..925V}. The broadening of the scattered points is likely due to elongated ionospheric structures \cite{mevius2016probing}.

The fitted frequency-dependent $r_{\rm diff}$ values are shown in Fig.~\ref{fig:rdiff}b. All three pointings show the expected increase of $r_{\rm diff}$ with observing frequency, consistent with the $\nu^{6/5}$ scaling derived from the first-order TEC term in Eq.~\ref{eq:tec_to_phase}. The largest diffractive scales are found for P139+79, indicating the weakest spatial ionospheric phase structure among the selected pointings. The smallest diffractive scales are found for P222+74, indicating the strongest spatial phase variations.

\subsection{Temporal ionospheric variability}

\begin{figure}[!ht]
    \centering
    \begin{minipage}{0.49\linewidth}
        \centering
        \includegraphics[width=\linewidth]{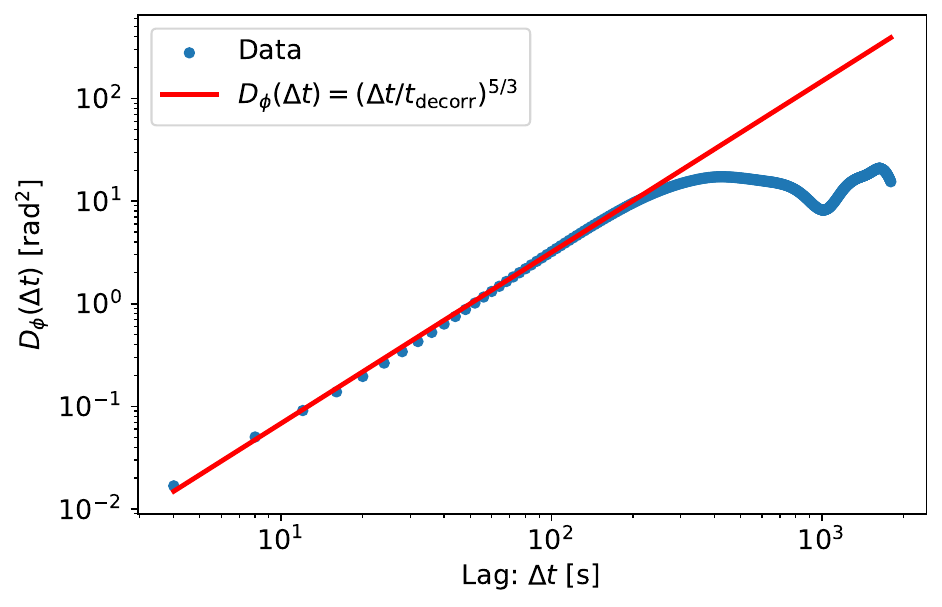}
        {(a) \small  Example temporal phase structure function $D_\phi(\Delta t)$ and Kolmogorov-like model for the first hour of field P139+79 at 30~MHz.}
    \end{minipage}
    \hfill
    \begin{minipage}{0.49\linewidth}
        \centering
        \includegraphics[width=\linewidth]{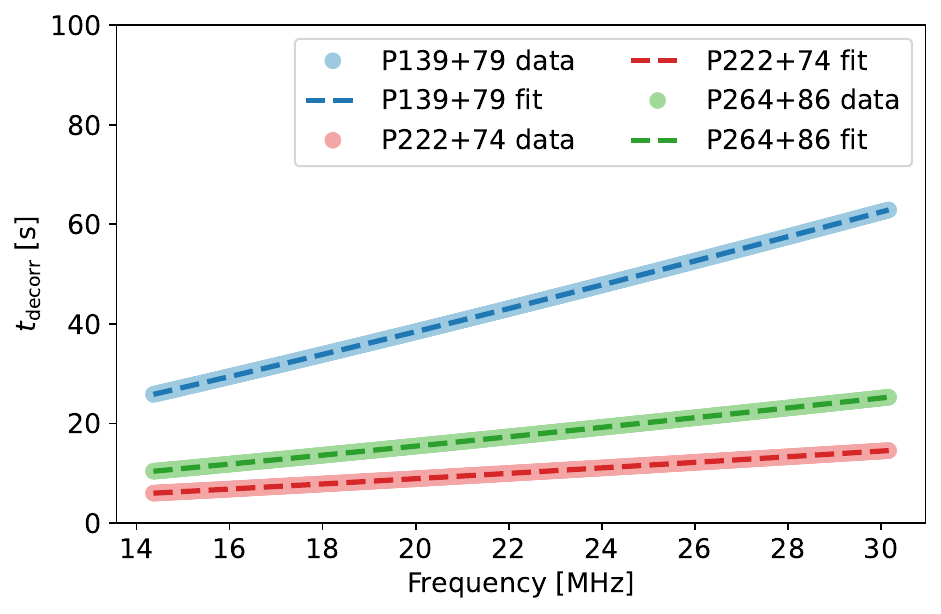}
        {(b) \small Frequency-dependent $t_{\rm decorr}$ values and $\nu^{6/5}$ fits for all three selected pointings.}
    \end{minipage}
    \vspace{0.1cm}
    \caption{Temporal ionospheric structure-function analysis and fitted decorrelation times.}
    \label{fig:tdiff}
\end{figure}

Figure~\ref{fig:tdiff} shows the corresponding temporal structure-function analysis. The example temporal structure function in Fig.~\ref{fig:tdiff}a rises at short time lags and then flattens at longer lags, which we interpret as the outer scale of ionospheric fluctuations as  described in Section~\ref{sec:iono_analysis}. 

The temporal decorrelation times also increase with frequency, broadly following the same $\nu^{6/5}$ behaviour expected if the temporal variations are produced by advection of a spatial phase screen. P139+79 has the longest $t_{\rm decorr}$ values, indicating the slowest temporal ionospheric variability. P222+74 has the shortest values, indicating the fastest phase variability.

\subsection{Angular scale for calibration transfer}

\begin{figure}[!ht]
    \centering
    \begin{minipage}{0.49\linewidth}
        \centering
        \includegraphics[width=\linewidth]{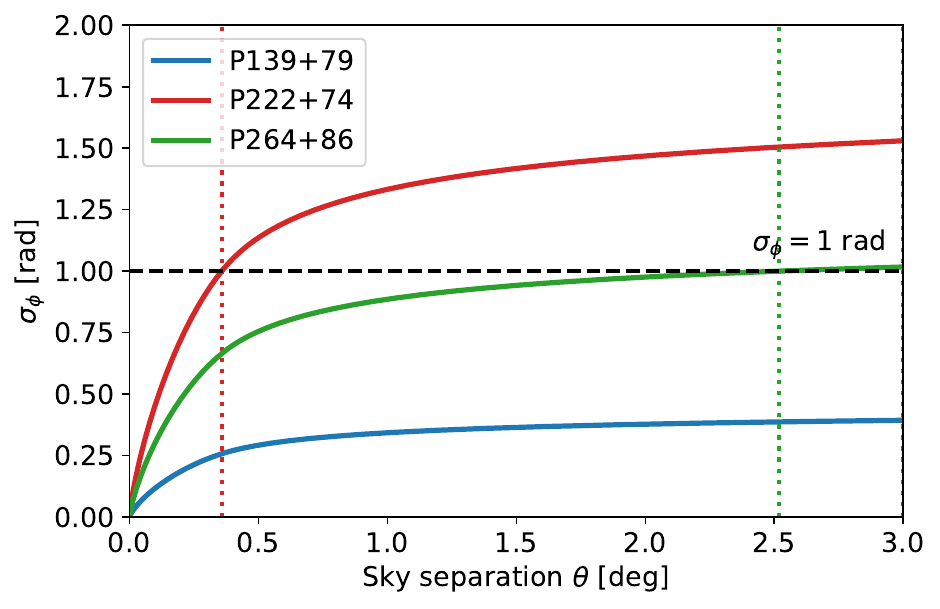}
        {(a) \small CS002LBA--CS302LBA}
    \end{minipage}
    \hfill
    \begin{minipage}{0.49\linewidth}
        \centering
        \includegraphics[width=\linewidth]{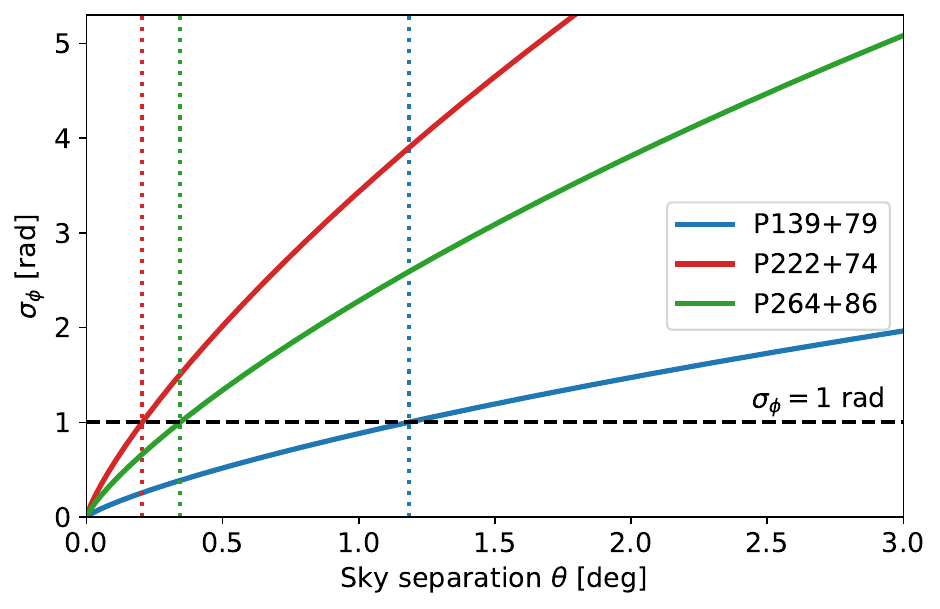}
        {(b) \small CS002LBA--RS509LBA}
    \end{minipage}
    \vspace{0.1cm}
    \caption{Expected RMS residual phase error as a function of sky separation for different baselines at 30~MHz.}
    \label{fig:theta}
\end{figure}

Figure~\ref{fig:theta} shows the expected RMS residual phase error as a function of angular separation on the sky at 30~MHz, computed using Eqs.~\ref{eq:sky_angle_variance}--\ref{eq:theta_1rad}. The horizontal dashed line marks the 1 rad threshold used to define $\theta_{\rm decorr}$. The two panels show the result for two different baselines, illustrating how the calibration-transfer scale depends on the station separation used in the four-point phase-screen geometry.

For the shorter baseline case, the RMS transfer error grows more slowly with sky separation, and several pointings remain below or near the 1 rad threshold across degree-scale separations. For the longer baseline case, the same ionospheric structure produces substantially larger transfer errors, and the 1 rad threshold is reached at smaller angular separations. 

Among the three pointings, P222+74 gives the largest transfer errors and therefore the smallest $\theta_{\rm decorr}$, while P139+79 gives the smallest transfer errors and the largest $\theta_{\rm decorr}$. The ordering is consistent with the relative spatial diffractive scales shown in Fig.~\ref{fig:rdiff}b.

\section{DISCUSSION}
\label{sec:discussion}

\begin{figure}[!ht]
    \centering
    \includegraphics[width=1\linewidth]{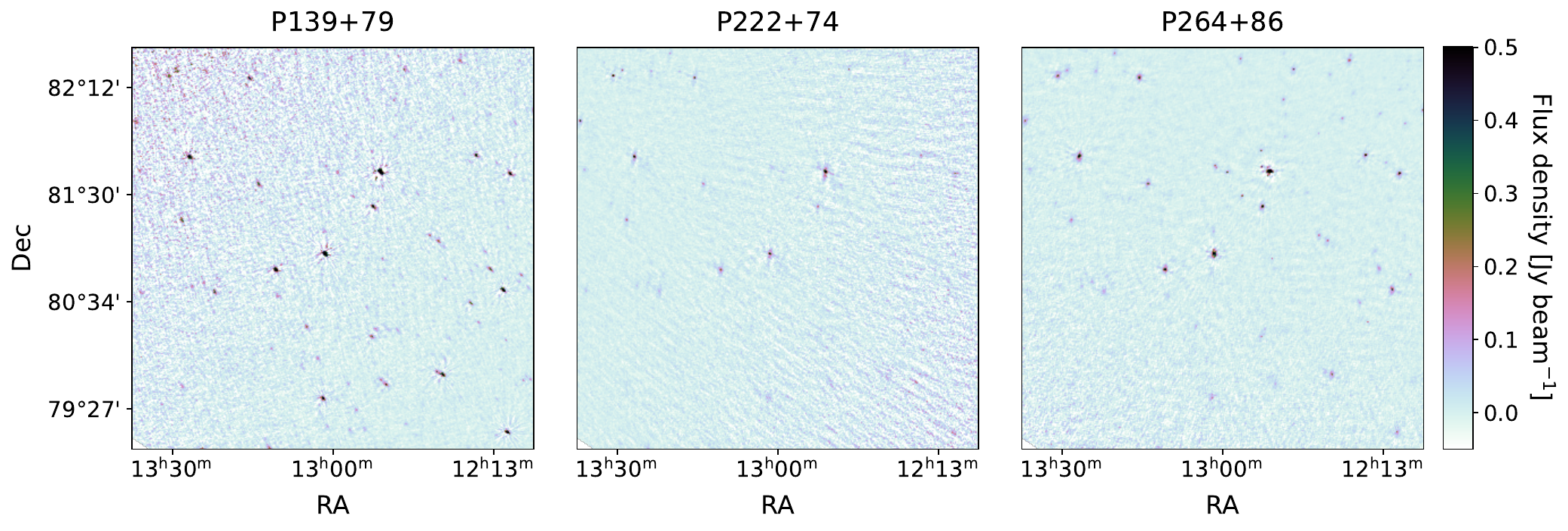}
    \caption{
    Primary-beam-corrected Stokes-I images of the common overlap region for the three selected pointings. The increased noise at the image edges is due to primary-beam correction.
    }
    \label{fig:image_comp}
\end{figure}

The ionospheric metrics derived in Sect.~\ref{sec:results} provide a direct way to interpret the final calibrated images. To test whether the differences in $r_{\rm diff}$, $t_{\rm decorr}$, and $\theta_{\rm decorr}$ are reflected in the image products, we compare the common sky region covered by the three pointings in Fig.~\ref{fig:image_comp}. The enhanced noise toward one side of each image is caused by primary-beam correction, because the overlap region lies close to the edge of at least one pointing.

The visual comparison follows the ordering expected from the ionospheric analysis. P139+79 shows the sharpest compact sources and the largest number of visible detections. P264+86 is intermediate: many sources remain detected, but they appear defocussed. P222+74 shows the poorest image quality, with several sources strongly defocussed or even sinking below the noise in the overlap region. This behaviour is consistent with Table~\ref{tab:iono_results}, where P139+79 has the largest diffractive scale and the largest angular decorrelation scale on the longest baselines, whereas P222+74 has the smallest values and is therefore expected to suffer the strongest residual phase errors after direction-dependent calibration.

\begin{figure}[!ht]
    \centering
    \includegraphics[width=0.7\linewidth]{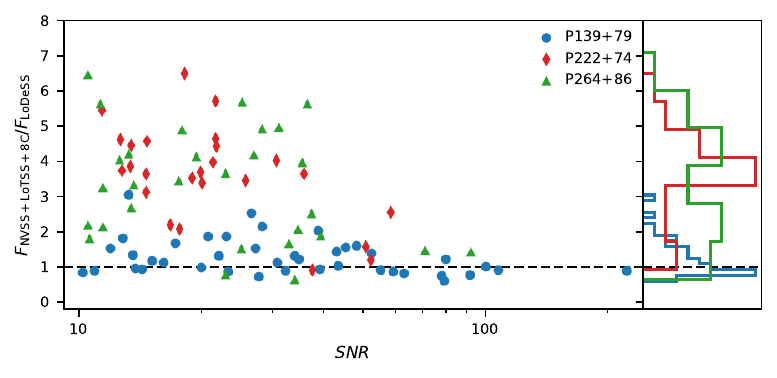}
    \caption{
    Ratio between the flux densities predicted at 23~MHz from external catalogues and the flux densities measured in the LoDeSS images. Different colours indicate the three pointings: P139+79 (blue), P222+74 (red), and P264+86 (green).
    }
    \label{fig:flux_comp}
\end{figure}

We quantified these differences by comparing the measured LoDeSS total flux densities with values predicted from existing radio catalogues. To avoid spurious detections and false matches between multi-component sources, we selected LoDeSS detections with ${\rm SNR}>10$ and required them to be isolated, with no neighbouring LoDeSS source within $90^{\prime\prime}$, approximately twice the synthesised beam. We applied a similar criterion to the third release of LOFAR Two-metre Sky Survey (LoTSS DR3\cite{shimwell2026lofar}) reference catalogue, requiring ${\rm SNR}>10$ and no neighbouring LoTSS source within $30^{\prime\prime}$. We then matched the selected LoDeSS sources to their nearest LoTSS counterparts within $60^{\prime\prime}$, keeping only one-to-one associations. This compact and isolated sample was finally cross-matched with the NRAO VLA Sky Survey (NVSS\cite{1998AJ....115.1693C}) and the 8C survey\cite{rees1990deep} to obtain flux-density measurements over a broad frequency range. For each source, we fit the catalogue flux densities as a function of frequency with a second-degree polynomial in log--log space. We used a second-degree polynomial to capture any intrinsic spectral curvature or low-frequency turnovers. The ratio between the flux density predicted by the fit at the LoDeSS central frequency of 23\,MHz and the measured LoDeSS flux densities is shown in Fig.~\ref{fig:flux_comp}.

The flux-density comparison gives the same ordering as the images. For P139+79, most sources lie close to the expected flux-density scale and show relatively small scatter. For P264+86, the scatter is substantially larger, indicating spatially variable residual calibration errors, although part of the sample still remains close to the expected scale. For P222+74, the measured LoDeSS flux densities are systematically lower than predicted and show large scatter. This is the expected behaviour of decorrelation, where residual phase errors reduce the coherent peak and integrated flux density recovered by the image, with the effect varying across the field depending on the distance to the nearest facet calibrator. 

What is the cause of the sources' flux drop from defocusing? One possibility is that the temporal cadence of calibration was insufficient to capture the ionospheric phase fluctuations. We used a calibration solution interval of 15~s for the stations that were farthest from the core of LOFAR. 
Only P222+74 has a decorrelation time comparable to, or slightly below, the shortest calibration interval. For P139+79 and P264+86, the temporal decorrelation times are longer than the relevant solution intervals. Therefore, although temporal under-sampling may contribute to the degradation of P222+74, it is unlikely to be the dominant cause of the field-to-field differences in the amount of defocusing.

Another possibility is that the size of a facet that shares the same calibration solution is larger than the diffractive scale and the corresponding angular decorrelation scale. When $r_{\rm diff}$ is small compared with the longest baselines used in the image, the residual phase difference between two nearby directions on the sky can become large. For P222+74 and P264+86, the angular scale over which calibration solutions can be transferred on the longest baselines is below half a degree. This is smaller than the typical angular scale covered by a single calibration facet, which is of the order of one degree. As a result, even after direction-dependent calibration, sources far from the facet calibrator will suffer residual phase errors that lead to defocusing as seen in Figs.~\ref{fig:image_comp} and~\ref{fig:flux_comp}. In contrast, P139+79 has a larger angular decorrelation scale than the typical size of facets. This leads to efficient calibration transfer to all sources in the facet and minimal defocusing. Based on these observations, we conclude that angular decorrelation is the dominant cause of ionospheric defocusing in our images. 

The most direct way to reduce the remaining ionospheric corruption is to increase the number of direction-dependent calibration facets, thereby reducing the angular distance over which each solution is transferred. Unfortunately, this is not feasible for the current survey data and processing setup. At decametre wavelengths, the background sky noise is high, and only a limited number of sources have enough signal-to-noise ratio to provide stable calibration solutions. Splitting the field into more directions would therefore reduce the calibrator signal-to-noise ratio per facet and could make the self-calibration unstable. 
The signal-to-noise ratio limitation can be overcome in two ways. (a) We can directly solve for the $\tau_1$ and $\tau_2$ ionospheric parameters during calibration rather than the current setup of solving for calibration phases channel by channel. (b) Solutions of neighbouring facets can be coupled by solving for an ionospheric model that is constrained to have a certain spatial decorrelation length.

A simple signal-to-noise scaling shows the possible gain from the first approach. If the calibration solution interval is kept fixed, the solution SNR scales approximately as ${\rm SNR} \propto \sqrt{\Delta \nu}$. Therefore, solving directly for $\tau_1$ and $\tau_2$ over the full 15 MHz observing bandwidth would increase the available calibration SNR by a factor of $\sqrt{15} \simeq 3.9$, compared with a 1 MHz effective bandwidth. Starting from the current 16-facet setup, this gain could, in principle, support approximately $16\sqrt{15} \simeq 60$ calibration directions at the same solution cadence.

Increasing the number of calibration directions would also reduce the typical angular size of each facet. Since the characteristic facet scale decreases approximately as $N_{\rm dir}^{-1/2}$, increasing the number of directions from 16 to about 60 would reduce the facet scale by a factor $\sqrt{16/60} \simeq 0.52$. Thus, full-band ionospheric fitting could reduce the typical facet size to roughly half of its present value. This would make the facet size closer to the angular decorrelation scale and should help in more perturbed fields, although it may still not be sufficient for the most strongly disturbed ionospheric conditions.

In the absence of these calibration improvements, the increased sensitivity expected from LOFAR~2.0 should also help, because more in-field sources will become usable as calibrators, allowing a denser facet layout.

A practical short-term mitigation is to restrict the maximum baseline length during imaging. Removing the longest baselines reduces the phase variance sampled by the array and increases the angular scale over which calibration solutions remain coherent. This would lower the angular resolution and moderately reduce the point-source sensitivity, but should help recover flux density in fields where the current long-baseline data are strongly decorrelated.

\section{CONCLUSION}
\label{sec:conclusion}

In this work, we have investigated how ionospheric conditions affect image quality in the LOFAR Decametre Sky Survey using three overlapping pointings observed under different ionospheric states. From the calibrator phase solutions, we derived three diagnostics of the ionospheric phase screen: the spatial diffractive scale, the temporal decorrelation time, and the angular decorrelation scale for transferring direction-dependent calibration solutions. These metrics were then compared with the final Stokes-I images and with the recovered flux-density scale.

We find that the spatial, temporal, and angular structure of the ionospheric phase is consistent with expectations from Kolmogorov turbulence. We also find that the temporal structure can be related to the spatial structure through characteristic advection speeds of around 100,m/s, as expected. Fields with a larger diffractive scale, a longer decorrelation time, and a larger angular scale for calibration transfer show less flux loss due to defocusing. Therefore, these ionospheric metrics, which can be readily computed from primary calibrator observations, can be used to predict image quality. Such a predictor could be incorporated into LOFAR's data quality calculator and potentially into a dynamic scheduler. Finally, the remaining ionospheric corruptions are primarily caused by the large angular sizes of the facets, which exceed the angular decorrelation scale for calibration transfer. These corruptions can be mitigated by exploiting the spectral and spatial coherence of the ionospheric phase during the calibration step.



\acknowledgments 
CMC and HKV acknowledge funding from the European Research Council via the starting grant `STORMCHASER' (grant number 101042416). This work made use of the Python packages \texttt{matplotlib} \cite{mpl} to generate the figures and \texttt{numpy} \cite{np} for computations. CMC acknowledges the use of ChatGPT for language editing.

\bibliography{report} 
\bibliographystyle{spiebib} 

\end{document}